# The Complexity of Kings[1]

*Edith Hemaspaandra*[2]
Department of Computer Science
Rochester Institute of Technology
Rochester, NY 14623 USA

*Lane A. Hemaspaandra*[3]
Department of Computer Science
University of Rochester
Rochester, NY 14627 USA

*Osamu Watanabe*[4]
Department of Mathematical and Computing Sciences
Tokyo Institute of Technology
Tokyo, 152-8552, Japan

June 13, 2005

---

[1] This report will appear as a technical/research report at U.R., T.I.T., and the Computing Research Repository.

[2] Supported in part by grant NSF-CCR-0311021. Work done in part while on sabbatical at the University of Rochester and while visiting the Tokyo Institute of Technology.

[3] Supported in part by grant NSF-CCF-0426761 and JSPS Invitational Fellowship S-05022. Work done in part while visiting the Tokyo Institute of Technology.

[4] Supported in part by "New Horizons in Computing" (2004–2006), an MEXT Grant-in-Aid for Scientific Research on Priority Areas.

**Abstract**

A king in a directed graph is a node from which each node in the graph can be reached via paths of length at most two. There is a broad literature on tournaments (completely oriented digraphs), and it has been known for more than half a century that all tournaments have at least one king [Lan53]. Recently, kings have proven useful in theoretical computer science, in particular in the study of the complexity of the semifeasible sets [HNP98,HT05] and in the study of the complexity of reachability problems [Tan01,NT02].

In this paper, we study the complexity of recognizing kings. For each succinctly specified family of tournaments, the king problem is known to belong to $\Pi_2^p$ [HOZZ]. We prove that this bound is optimal: We construct a succinctly specified tournament family whose king problem is $\Pi_2^p$-complete. It follows easily from our proof approach that the problem of testing kingship in succinctly specified graphs (which need not be tournaments) is $\Pi_2^p$-complete. We also obtain $\Pi_2^p$-completeness results for $k$-kings in succinctly specified $j$-partite tournaments, $k, j \geq 2$, and we generalize our main construction to show that $\Pi_2^p$-completeness holds for testing $k$-kingship in succinctly specified families of tournaments for all $k \geq 2$.

# 1 Introduction

## 1.1 General Introduction and Overview

We study the complexity of recognizing kings.

Throughout this paper, unless otherwise stated, $\Sigma = \{0, 1\}$, and each graph will be simple (no self-loops) and directed, and will have at least one node. A node $v$ of a graph $G = (V_G, E_G)$ is said to be a *king* exactly if each node in $G$ can be reached from $v$ via a path of length at most two. In the 1950s, Landau noted the simple but lovely result that every *tournament*—i.e., every graph $G$ such that for each pair of distinct nodes $a, b \in V_G$, exactly one of the directed edges $a \to b$ and $b \to a$ occurs in $E_G$—contains at least one king [Lan53]. (Throughout this paper "$a \to b$ occurs in $E$" and "$(a, b) \in E$" will be synonymous. In the former form, we will sometimes simply write "$a \to b$" when the edge set $E$ is clear from context.)

When tournaments are specified explicitly in the natural way, it is not hard to see that the king problem is first-order definable, and so by Lindell [Lin92] is in $\text{AC}^0$ (see [Tan01, NT02], which have analogous discussions for reachability—and which involve kings in their proofs; we also mention that a number of papers interestingly study king-respecting sequencing in nonsuccinctly specified tournaments, see, e.g., [SSW03, HC03] and the references therein). Thus, in this paper we focus mostly on the case that actually arises in the study of the semifeasible sets, namely, succinctly specified tournament families, though we will also resolve the general case of succinctly specified graphs (which is in some sense an easier result, since the lower bound is the challenging part).

In particular, we focus on families of tournaments that are defined, uniformly, by a P-time function. To formalize this, we in effect adopt the existing formalism (though, for clarity, not the naming scheme) that is already provided by the theory of semifeasible sets. In particular, we say a function $f$ is a *tournament family specifier* exactly if

1. $f$ is a polynomial-time computable function.
2. $(\forall x, y \in \Sigma^*)[f(x, y) = f(y, x)]$.
3. $(\forall x, y \in \Sigma^*)[f(x, y) = x \vee f(x, y) = y]$.

We interpret this as specifying, in the following way, a family of tournaments, one per length. At each length $n$, the nodes in the length $n$ tournament specified by $f$ will be the strings in $\Sigma^n$. For each two distinct nodes among these, $x$ and $y$, the edge between them will go from $x$ to $y$ if $f(x, y) = x$ and will go from $y$ to $x$ if $f(x, y) = y$. Since our function $f$ always chooses one of its inputs and is commutative, this indeed yields a family of tournaments. We will call the tournament just described *the length $n$ tournament induced by $f$*. This formalism is precisely the one that plays an important role in the study of the semifeasible sets, and in Section 1.2 we will explain what the connection is, and why we were motivated to study the complexity of kings.[1]

---

[1]The reader may note that the constraints in the definition of tournament family specifiers apply even between strings of different lengths, and yet this is never used in our proofs of results about tournament

For $f$ a tournament family specifier, the set whose complexity we will first study is

$$\text{Kings}_f = \{x \mid x \text{ is a king in the length } |x| \text{ tournament induced by } f\}.$$

$\Pi_2^p = \text{coNP}^{\text{NP}}$ is the "$\Pi$-side" second level of the polynomial hierarchy [MS72,Sto76]. It is already known that for each tournament family specifier $f$, $\text{Kings}_f \in \Pi_2^p$ [HOZZ]. The central result of this paper shows that that result is optimal: We prove that there is a tournament family specifier $f$ such that $\text{Kings}_f$ is $\Pi_2^p$-complete (i.e., is $\leq_m^p$-complete for $\Pi_2^p$). We will note that if one changes one's notion of tournament family specifier to an analogous one that specifies an infinite family of graphs, $\Pi_2^p$-completeness still holds. (Since the difficult part here is the $\Pi_2^p$-hardness lower bound, proving our result for the special case of tournaments is in fact harder than proving it in the general case. But we undertake the greater challenge of tournaments both because it yields easily the other cases, and because it better connects to the motivating issue from the theory of semifeasible sets.)

It is easy to see from our proof that we also obtain $\Pi_2^p$-completeness for the following problem, which some may find more natural as it deals not with uniform families of problems but just with individual inputs: "Given a succinctly specified (via a circuit following the Galperin–Wigderson model[2]) graph $G$ and a node $x \in V_G$, is $x$ a king of $G$?" Note that we do not require that $G$ be a tournament.

Recall that the central result of this paper is that there is a tournament family specifier $f$ such that $\text{Kings}_f$ is $\Pi_2^p$-complete. Can one in some broad cases do better than $\Pi_2^p$-completeness? We note that for tournament family specifiers that are associative, the king problem in fact is always in coNP, yet for associative specifiers we also prove that the king problem cannot be coNP-complete unless P = NP. On the other hand, we show that various natural complexity levels are precisely the complexity of the king problem of some tournament family specifiers: There are tournament family specifiers $f$ for which $\text{Kings}_f$ is coNP-complete, and there are tournament family specifiers $f$ for which $\text{Kings}_f$ is NP-complete.

The results mentioned above, about the complexity of kings in tournaments (and graphs), appear in Section 2. Recall that kings are nodes that cover the whole graph via paths of length at most two. The notion has been generalized as follows. For each fixed $k$, a $k$-king in a digraph is a vertex that can reach all other vertices in the digraph via

---

family specifiers since specifiers specify different, separate tournaments at each length. This observation is correct. The reason we have required the constraints to hold globally is simply because our motivating notion, P-selectors from P-selectivity theory, has these constraints holding globally. However, if one changed the definition of tournament family selector (and, later, graph family specifier and associative tournament family specifier) to simply require the various constraints to hold merely between strings $x$ and $y$ of the same length, then every one of our results that uses the words "family specifier" in its theorem statement would still hold. In particular, the $\Pi_2^p$-completeness, coNP-completeness, and NP-completeness results would still hold.

[2]The Galperin–Wigderson model [GW83] for succinctly specifying a graph (simple and directed, since in this paper all graphs are taken to be simple and directed) is that one gives a circuit with $2n$ inputs, $x_1, \ldots, x_n, y_1, \ldots, y_n$, and it specifies a graph on the nodes $\Sigma^n$ as follows: There are no self-loops, and for each $x \neq y$, $|x| = |y| = n$, $x \to y$ is an edge of $G$ exactly if the circuit on the $2n$ input bits $x \cdot y$ evaluates to 1, where $\cdot$ denotes concatenation.

paths of length at most $k$. The existence of $k$-kings has been intensely studied, especially with respect to *multipartite* tournaments, e.g., it is known that no multipartite tournament that has two or more nodes of indegree zero can have a $k$-king for any $k$, and it is known that every multipartite tournament having at most one vertex of indegree zero does have a 4-king ([Gut86,PT91], see also [BG98] and the references therein).

Section 3 studies the complexity of testing $k$-kingship in tournaments (and graphs) for values of $k$ other than the $k = 2$ case handled in Section 2. We will show that for $k = 1$ the problem is simpler, but for $k > 2$ it remains $\Pi_2^p$-complete, via a proof that rather interestingly is not a perfect analog of the $k = 2$ case.

Section 4 studies the complexity of testing $k$-kingship in multipartite tournaments, which as mentioned above is the domain in which $k$-kingship, $k > 2$, has primarily been studied previously (though to the best of our knowledge never regarding the computational complexity of recognizing $k$-kings). This section is in the model of studying succinct graphs (not graph families via a specifier), which seems the most natural model for that study. We completely capture the complexity in each case as either belonging to P or being $\Pi_2^p$-complete.

Section 5 presents some open issues for further study.

## 1.2 Connections to Semifeasible Sets

Selman initiated the study of the semifeasible sets in a series of papers starting in 1979 [Sel79,Sel81,Sel82b,Sel82a]. A set $A$ is *P-selective* (or *semifeasible*) exactly if there is a polynomial-time computable function $f$ (called a *P-selector function for $A$*) such that, for each $x, y \in \Sigma^*$, (a) $f(x, y) = x$ or $f(x, y) = y$, and (b) $\{x, y\} \cap A \neq \emptyset \implies f(x, y) \in A$. (For a discussion of the motivation for, and examples and applications of, the semifeasible sets, see [HT03], especially [HT03, Preface].) It was soon observed by Ko [Ko83] that whenever there exists such a function for $A$, then there exists such a function that in addition is commutative, i.e., $(\forall x, y)[f(x, y) = f(y, x)]$. However, note that (aside from the connection with $A$) such functions are precisely tournament family specifier functions. Indeed, this connection between commutative P-selector functions and families of tournaments has been very useful in obtaining results about the P-selective sets. More particularly, this connection has been used—starting with the important work of Ko [Ko83] establishing that all P-selective sets have small circuits—to get results on the *advice complexity* of the P-selective sets.

In the rest of this subsection, we will speak a bit about advice classes. So we now quickly define those, although we assume the reader is generally familiar with the notion. However, readers who are interested just in our completeness results, or who are unfamiliar with the notion of advice classes, can safely skip to Section 1.3.

Advice classes, introduced by Karp and Lipton [KL80], ask what class of sets a given complexity class can accept when given a small amount of extra information that depends only on the length of the string whose membership is being asked about. These classes have been the subject of extensive study, and are key tools in showing that certain complexity assumptions would collapse the polynomial hierarchy. The classes are formally defined as follows (we copy this definition from [HT03], but it is faithful to the exact notion of Karp and Lipton).

**Definition 1.1**  1. Let $f : \mathbb{N} \to \mathbb{N}$ be any function. Let $\mathcal{C}$ be any collection of sets. Define

$$\mathcal{C}/f = \{A \,|\, (\exists B \in \mathcal{C})\, (\exists h : \mathbb{N} \to \Sigma^*)\, [(\forall n)\, [|h(n)| = f(n)] \\ \land\, (\forall x \in \Sigma^*)\, [x \in A \iff \langle x, h(|x|)\rangle \in B]]\}.$$

2. Let $\mathcal{F}$ be any class of functions mapping from $\mathbb{N}$ to $\mathbb{N}$. Define

$$\mathcal{C}/\mathcal{F} = \{A \,|\, (\exists f \in \mathcal{F})\, [A \in \mathcal{C}/f]\}.$$

3. "linear" ("poly") will denote all functions from $\mathbb{N}$ to $\mathbb{N}$ where the value of the output is linearly bounded (polynomially bounded) in the value of the input.

For example, P/$n$ is the class of sets that are, informally put, so simple that with just $n$ extra bits of help at each length $n$ they can be accepted by deterministic polynomial-time machines. And Ko's [Ko83] result mentioned earlier can be stated as $\{L \,|\, L$ is a P-selective set$\} \subseteq$ P/poly.

Let us return to the P-selective sets. When seeking to prove an advice result for a P-selective set, the standard approach is to take a commutative P-selector function for the set and view it as a tournament family specifier function, and then to exploit some properties of the tournament to construct algorithms that show that the set can be accepted with surprisingly little advice. (Sometimes the focus of the arguments is not on the entire tournament at a given length but rather is on the subtournament consisting of just the strings of that length that happen to belong to the P-selective set. However, that is not of crucial interest to us. And in the key motivating example, the focus in fact is on all the strings of a given length.)

Due to this close connection between advice results and tournaments, it is hardly surprising that Landau's result that all tournaments (by which we always mean all nonempty tournaments) have at least one king has proven useful in the study of P-selective sets and their advice. For example, Landau's result underpins the proof that all P-selective sets belong not merely to NP/linear [HT96] but even to NP$_1$/linear [HNP98], where NP$_1$ is the class of all sets accepted by NP machines using only linearly many nondeterministic guess bits (this class forms the first level of the limited nondeterminism hierarchy of [KF77], cf. [KF80]).

The particular use of Landau's result that brought our interest to this topic is related but somewhat different. As mentioned earlier, it has recently been noted (this is clear by brute force) that for any tournament family specifier $f$ the problem Kings$_f$ belongs to $\Pi_2^p$. However, note that in a tournament, induced by any commutative P-selector function for a P-selective set $A$ (in our terminology, the P-selector function serves as an appropriate tournament family specifier), we have that for each length $n$ it holds that: If any string belongs to $A$ at length $n$, then all kings in the tournament at length $n$ belong to $A$. And by Landau's result, there always is at least one king in each such tournament. This led Hemaspaandra and Torenvliet to the following observation.

**Theorem 1.2 ([HT05])** *The class of* P*-selective sets is not* $\Pi_2^p/1$*-immune. (That is, each infinite* P*-selective set has an infinite subset belonging to* $\Pi_2^p/1$*.)*

A natural way to seek to improve that nonimmunity upper bound would be to improve the $\Pi_2^p$ upper bound on the complexity of $\text{Kings}_f$. However, the core result of this paper precisely shows that that particular line of attack is essentially hopeless, since we prove that there exists a tournament family specifier $f$ such that $\text{Kings}_f$ is $\Pi_2^p$-complete. It however remains conceptually possible that some other attack might lower the $\Pi_2^p/1$ bound. Also, as mentioned earlier, the core result of this paper directly shows that the $\Pi_2^p$ upper bound of [HOZZ] has a matching lower bound.

### 1.3 Comparison with Other Work

We now mention another paper that exploits Landau's king result, and we compare and contrast that other paper to the work of Section 2 of the current paper. Tantau ([Tan01], see also the more general conference version by Nickelsen and Tantau [NT02]) critically uses Landau's king result in proving that nonsuccinct tournament reachability is first-order definable. That same paper proves (not via the king result) that in the Galperin–Wigderson model succinct tournament reachability is $\Pi_2^p$-complete. This is a lovely result, but differs from our core result (i.e., our Theorem 2.1: There is a tournament family specifier $f$ such that $\text{Kings}_f$ is $\Pi_2^p$-complete) in multiple ways.

First, for Tantau, the $\Pi_2^p$ upper bound is critically dependent on the graphs being tournaments. In contrast, for us and the king problem, the $\Pi_2^p$ upper bound easily holds for both tournaments and general graphs, and for each of those, in both the family-specifier model and in the Galperin–Wigderson model of inputting individual graph instances as a circuit.

Second, though this is less a difference than a caution, it is true that a node $v$ is a king exactly if for each other node $w$ one can reach $w$ from $v$ via some path of length *at most two*. But that connection in no way implies that hardness results for reachability (or even reachability via paths of length at most two) imply hardness results for the king problem in our model.

Third, we stress that our core result is in the model where we study a tournament family specifier. In this model, for the tournament family specifier $f$, there is just one tournament at each length. Thus, one might perhaps expect merely collective $\leq_m^p$-hardness for $\text{TALLY} \cap \Pi_2^p$ (i.e., one might expect merely that $\text{TALLY} \cap \Pi_2^p \subseteq \{L \mid (\exists \text{ tournament family specifier } f)[L \leq_m^p \text{Kings}_f]\}$), rather than having outright $\Pi_2^p$-hardness hold. In fact, in our most central proof—that of Theorem 2.1—one main obstacle is to show how an exponential number of separate kingship problems can be embedded all into a single tournament in such a way that they do not cross-pollute each other. In contrast, in the Galperin–Wigderson model this difficult issue does not exist, as there the input is a (node whose kingship we are interested in, and a succinctly specified) single graph or tournament.

Why do we in Theorem 2.1 seek to clear this higher bar of studying families? Because we find it interesting, because our proof will be so general as to in effect give all the other cases easily, and because this is the model of the existing $\Pi_2^p$ upper bound of [HOZZ] and so, to show that that upper bound cannot be improved, we cannot validly cheat by changing to an easier-to-resolve issue.

On the other hand, Tantau's [Tan01] (and Nickelsen–Tantau's [NT02]) work does share with this paper one important property. Both their work and ours—in contrast with the many earlier papers showing that a wide variety of properties of succinctly specified graphs are PSPACE-hard [Wag84,Wag86,PY86]—pinpoints succinctly specified graph problems whose complexity falls at a substantially lower level, namely, $\Pi_2^p$ (see also [GW83,Wag86]).

Finally, we mention that the generalization achieved in the transition from Tantau's paper [Tan01] to the Nickelsen–Tantau paper [NT02], namely, moving from tournaments to graphs of bounded independence number, is not particularly interesting in our case. For the king problem in our setting, our $\Pi_2^p$-hardness lower bounds are immediately inherited by the more flexible case of independence numbers beyond two (i.e., beyond the number possessed by tournaments), and though the $\Pi_2^p$ upper bound doesn't automatically transfer, it is for those problems in our case immediately obvious, directly, that it holds.

## 2 The Complexity of Kingship (i.e., 2-Kingship) in Tournaments and Graphs

We now prove our core result.

**Theorem 2.1** *There is a tournament family specifier $f$ such that $\mathrm{Kings}_f$ is $\Pi_2^p$-complete.*

The proof has two parts. First, we show how a single $\Pi_2^p$-type formula can be converted to a king problem. Then—since $\mathrm{Kings}_f$ speaks of just one tournament per length—we show how one can in effect embed an exponential number of king problems into a single length without creating any damaging cross-pollution.

**Proof of Theorem 2.1** It is easy to see that for every tournament family specifier $f$, $\mathrm{Kings}_f$ is in $\Pi_2^p$ [HOZZ]. We will in this proof define a tournament family specifier $f$ such that $\mathrm{Kings}_f$ is $\Pi_2^p$-hard. For this tournament family specifier $f$, we will show $\Pi_2^p$-hardness for $\mathrm{Kings}_f$ by a reduction from $\forall\exists$SAT, the set of true fully quantified Boolean formulas where all universal quantifiers precede all existential quantifiers. It is well known that $\forall\exists$SAT is $\Pi_2^p$-complete [SM73,Wra76]. Without loss of generality, we will assume that all formulas in $\forall\exists$SAT have the same number of universally quantified variables as existentially quantified variables and that the number of universally quantified variables is greater than 0. We will call formulas of the right form $\forall\exists$-formulas, i.e., we say that $\phi$ is an $\forall\exists$-formula if and only if there exists an integer $n > 0$ and a propositional formula $\phi'$ such that $\phi = \forall x_1 \cdots \forall x_n \exists y_1 \cdots \exists y_n \phi'(x_1, \ldots, x_n, y_1, \ldots, y_n)$. Using this terminology, for us $\forall\exists$SAT will denote the set of true $\forall\exists$-formulas.

We will define our tournament family specifier $f$ in two stages. First we will define for every $\forall\exists$-formula $\phi$ a tournament $T_\phi = (V_\phi, E_\phi)$ such that a specific "'potential king" node in $V_\phi$ is a king in $T_\phi$ if and only if $\phi$ is true. We will then show how to combine all these tournaments into one family of tournaments (specified by $f$) without disturbing the king-ness of the potential king in $T_\phi$ for all $\forall\exists$-formulas $\phi$.

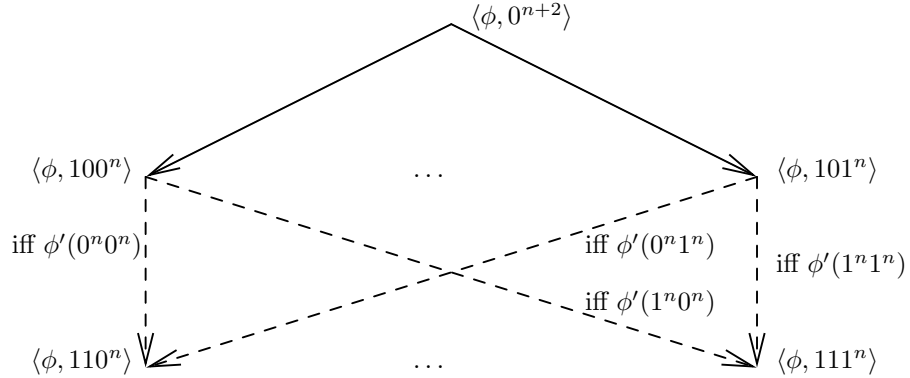

Figure 1: First stage of the 2-Kings$_f$ $\Pi_2^p$-hardness construction

To make sure that all tournaments $T_\phi$ can be combined in the desired way, we will use a binary pairing function $\langle \cdot, \cdot \rangle$ that is polynomial-time computable and polynomial-time invertible (by the latter we mean that the range of the function is in P, and that there exist two polynomial-time computable functions that, given a string in the range of the function, return the first and the second component, respectively) such that for all strings $x, x', y, y' \in \Sigma^*$, if $|x| = |x'|$ and $|y| = |y|$ then $|\langle x, y \rangle| = |\langle x', y' \rangle|$ and such that for all $x, y \in \Sigma^*$, $\langle x, y \rangle \notin 0^*$. It is easy to see that such a pairing function exists, for example by defining $\langle x_1 x_2 \cdots x_n, y \rangle$ as $0x_1 0x_2 \cdots 0x_n 1y$ for all $x_1, x_2, \ldots, x_n \in \Sigma$ and $y \in \Sigma^*$.

Let $\phi$ be an $\forall\exists$-formula. Let $n > 0$ and let $\phi'$ be a propositional formula such that $\phi = \forall x_1 \cdots \forall x_n \exists y_1 \cdots \exists y_n \phi'(x_1, \ldots, x_n, y_1, \ldots, y_n)$. We will define tournament $T_\phi = (V_\phi, E_\phi)$ in such a way that $\langle \phi, 0^{n+2} \rangle$ is a king in $T_\phi$ if and only if $\phi$ is true.

Figure 1 gives a pictorial representation of $T_\phi$. The nodes in $V_\phi$ are arranged in three layers. The first layer consists of the potential king $\langle \phi, 0^{n+2} \rangle$. The second layer contains a node $\langle \phi, 10y \rangle$ for every $y \in \Sigma^n$. These $2^n$ nodes correspond to the $2^n$ possible assignments to the $y$-variables in $\phi'$. The third layer contains a node $\langle \phi, 11x \rangle$ for every $x \in \Sigma^n$. These $2^n$ nodes correspond to the $2^n$ possible assignments to the $x$-variables in $\phi'$. In the figure, we use the convention that missing edges between nodes at different levels go "up," and that missing edges between nodes at the same level go "right."

Formally, $T_\phi = (V_\phi, E_\phi)$ is defined as follows. $V_\phi = \{\langle \phi, 0^{n+2} \rangle\} \cup \{\langle \phi, 10y \rangle \mid y \in \Sigma^n\} \cup \{\langle \phi, 11x \rangle \mid x \in \Sigma^n\}$. Note that all strings in $V_\phi$ have the same length by the properties of the pairing function. (There will be strings in $\Sigma^*$ at that length that are not in $V_\phi$; this will be handled in the second stage of our construction.) For all $z, z' \in \Sigma^{n+2}$ such that $\langle \phi, z \rangle, \langle \phi, z' \rangle \in V_\phi$ and $z < z'$ (i.e., $z <_{\text{lexicographic}} z'$), let $(\langle \phi, z \rangle, \langle \phi, z' \rangle) \in E_\phi$ if and only if

- ($z = 0^{n+2}$ and $z' = 10y$ for some $y \in \Sigma^n$), or

- ($z = 10y$ for some $y \in \Sigma^n$ and $z' = 11x$ for some $x \in \Sigma^n$ and $\phi'(xy)$), or
- ($z = 10y$ for some $y \in \Sigma^n$ and $z' = 10y'$ for some $y' \in \Sigma^n$), or
- ($z = 11x$ for some $x \in \Sigma^n$ and $z' = 11x'$ for some $x' \in \Sigma^n$).

For all $z, z' \in \Sigma^{n+2}$ such that $\langle \phi, z \rangle, \langle \phi, z' \rangle \in V_\phi$ and $z > z'$, let $(\langle \phi, z \rangle, \langle \phi, z' \rangle) \in E_\phi$ if and only if $(\langle \phi, z' \rangle, \langle \phi, z \rangle) \notin E_\phi$.

It is clear that $T_\phi$ is a tournament. It is immediate from the construction that

- $\langle \phi, 0^{n+2} \rangle$ is a king in tournament $T_\phi$ if and only if for all $x \in \Sigma^n$, $\langle \phi, 0^{n+2} \rangle$ reaches $\langle \phi, 11x \rangle$ in two steps,

- for all $x \in \Sigma^n$, $\langle \phi, 0^{n+2} \rangle$ reaches $\langle \phi, 11x \rangle$ in two steps if and only if there exists a $y \in \Sigma^n$ such that $(\langle \phi, 10y \rangle, \langle \phi, 11x \rangle) \in E_\phi$, and

- for all $x, y \in \Sigma^n$, $(\langle \phi, 10y \rangle, \langle \phi, 11x \rangle) \in E_\phi$ if and only if $\phi'(xy)$.

From these observations, the following claim follows immediately.

**Claim 2.2** $\langle \phi, 0^{n+2} \rangle$ *is a king in tournament $T_\phi$ if and only if $\phi$ is true.*

We will now show how to combine the $T_\phi$ tournaments into a family of tournaments specified by $f$ in such a way that for every $\forall \exists$-formula $\phi$, $\langle \phi, 0^{n_\phi+2} \rangle$ is a king in $T_\phi$ if only if $\langle \phi, 0^{n_\phi+2} \rangle \in \text{Kings}_f$, where $n_\phi$ is the number of universally quantified variables in $\phi$.

For every $\forall \exists$-formula $\phi$ and for all $x, y \in V_\phi$, let $f(x, y) = x$ if and only if $(x, y) \in E_\phi$. Note that by definition of $\langle \cdot, \cdot \rangle$, all elements in $V_\phi$ have the same length. We need to ensure that the rest of $f$ is specified in such a way as to not disturb the king-ness of $\langle \phi, 0^{n+2} \rangle$. Note that it is unavoidable that there exist different $\forall \exists$-formulas $\phi$ and $\psi$ such that strings in $V_\phi$ and $V_\psi$ have the same length.

Figure 2 gives a pictorial representation of of the tournament induced by $f$ at length $m$. In the figure, $\phi_1, \phi_2, \ldots, \phi_k$ are all $\forall \exists$-formulas such that $V_{\phi_i} \subseteq \Sigma^m$. The $\phi_i$'s are ordered lexicographically, in ascending order. For readability, we write $T_i$ for $T_{\phi_i}$ and $n_i$ for $n_{\phi_i}$. Note that for all formulas $\phi$, $0^m \notin V_\phi$ (by properties of the pairing function) and $\langle \phi, 010^{n_\phi} \rangle \notin V_\phi$. We use the convention that all missing arrows between $T_i$ and $T_j$ go right, that all missing arrows between $\langle \phi_i, 010^{n_i} \rangle$ and $\langle \phi_j, 010^{n_j} \rangle$ go right, that all missing arrows between nodes in "all other strings" go from lexicographically smaller strings to lexicographically larger strings, and that all other missing arrows go up. This completely specifies the tournament on strings of length $m$.

Formally, we define $f$ as follows. Let

$$\text{Other} = \Sigma^* - \left( 0^* \cup \{\langle \phi, 010^{n_\phi} \rangle \mid \phi \text{ is a } \forall \exists \text{-formula}\} \cup \bigcup \{V_\phi \mid \phi \text{ is a } \forall \exists \text{-formula}\} \right).$$

The set *Other* is clearly in P, since the pairing function is polynomial-time invertible. For all $m$ and all $z, z' \in \Sigma^m$, let $f(z, z') = z$ if and only if

- $z = z'$, or

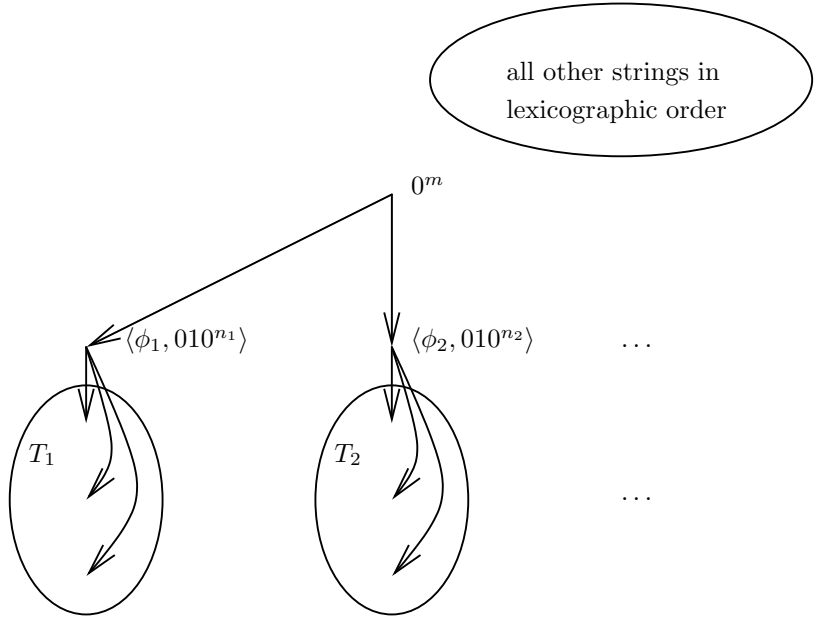

Figure 2: Second stage of the 2-Kings$_f$ $\Pi_2^p$-hardness construction

- $z = 0^m$ and $z' = \langle \phi, 010^{n_\phi} \rangle$ for some $\forall\exists$-formula $\phi$, or

- $z = 0^m$ and $z' \in \textit{Other}$, or

- $z = \langle \phi, 010^{n_\phi} \rangle$ for some $\forall\exists$-formula $\phi$ and $z' = \langle \psi, 010^{n_\psi} \rangle$ for some $\forall\exists$-formula $\psi$ and $\phi < \psi$, or

- $z = \langle \phi, 010^{n_\phi} \rangle$ for some $\forall\exists$-formula $\phi$ and $z' \in V_\phi$, or

- $z = \langle \phi, 010^{n_\phi} \rangle$ for some $\forall\exists$-formula $\phi$ and $z' \in \textit{Other}$, or

- $z \in V_\phi$ for some $\forall\exists$-formula $\phi$ and ($z' = 0^m$ or $z' \in \textit{Other}$), or

- $z \in V_\phi$ for some $\forall\exists$-formula $\phi$ and $z' = \langle \psi, 010^{n_\psi} \rangle$ for some $\forall\exists$-formula $\psi$ and $\phi \neq \psi$, or

- $z \in V_\phi$ for some $\forall\exists$-formula $\phi$ and $z' \in V_\phi$ and $(z, z') \in E_\phi$, or

- $z \in V_\phi$ for some $\forall\exists$-formula $\phi$ and $z' \in V_\psi$ for some $\forall\exists$-formula $\psi$ and $\phi < \psi$, or

- $z, z' \in \textit{Other}$ and $z < z'$.

For all $z, z' \in \Sigma^m$, if $f(z, z') \neq z$, then let $f(z, z') = z'$. The definition of $f$ on strings of different lengths is irrelevant (as long as $f$ remains a tournament family specifier). To be complete, we define $f(z, z') = z$ if $|z| < |z'|$ and $f(z, z') = z'$ if $|z| > |z'|$.

It is immediate that for all $x, y \in \Sigma^*$, $f(x,y) = f(y,x)$ and that $f(x,y) = x$ or $f(x,y) = y$. Since the pairing function is polynomial-time invertible, it is easy to see that $f$ is computable in polynomial time. Thus $f$ is a tournament family specifier.

To finish the proof, let $\phi$ be a $\forall\exists$-formula and let $n = n_\phi$. It remains to show that $\langle \phi, 0^{n+2} \rangle \in \text{Kings}_f$ if and only if $\phi$ is true. By Claim 2.2, it suffices to show that $\langle \phi, 0^{n+2} \rangle \in \text{Kings}_f$ if and only if $\langle \phi, 0^{n+2} \rangle$ is a king in $T_\phi$. Let $m = |\langle \phi, 0^{n+2} \rangle|$.

First suppose that $\langle \phi, 0^{n+2} \rangle$ is a king in $T_\phi$. It is easy to see from the definition of $f$ at length $m$ that $\langle \phi, 0^{n+2} \rangle$ reaches all strings in $\Sigma^m - V_\phi$ in one or two steps. (Strings of length $m$ that are in $V_\psi$ for some $\forall\exists$-formula $\psi \neq \phi$ are reached in two steps via $\langle \psi, 010^{n_\psi} \rangle$.) It follows that $\langle \phi, 0^{n+2} \rangle$ is a king in the tournament induced by $f$ on strings of length $m$, i.e., $\langle \phi, 0^{n+2} \rangle \in \text{Kings}_f$.

For the converse, suppose that $\langle \phi, 0^{n+2} \rangle$ is a king in the tournament induced by $f$ on strings of length $m$. Then every string of length $m$ can be reached from $\langle \phi, 0^{n+2} \rangle$ in at most two steps. In particular, every element from $V_\phi$ can be reached from $\langle \phi, 0^{n+2} \rangle$ in at most two steps. By the construction of $f$, there do not exist length two paths $v \to w \to v'$ such that $v, v' \in V_\phi$ and $w \notin V_\phi$ in the tournament induced by $f$ on length $m$. This implies that every node in $V_\phi$ can be reached from $\langle \phi, 0^{n+2} \rangle$ by a path of length at most two such that all nodes of the path are in $V_\phi$. It follows that $\langle \phi, 0^{n+2} \rangle$ is a king in the tournament induced by $f$ on $V_\phi$, and thus $\langle \phi, 0^{n+2} \rangle$ is a king in $T_\phi$.

To be explicit, our $\leq_m^p$-reduction $g$ from $\forall\exists\text{SAT}$ to $\text{Kings}_f$ is as follows. Note that $Other \neq \emptyset$ and no element of $Other$ belongs to $\text{Kings}_f$. Let $out$ be any fixed element of $Other$. Our reduction $g$ on an arbitrary input $\phi$ will output $out$ if $\phi$ is not a $\forall\exists$-formula, and otherwise will output $\langle \phi, 0^{n_\phi+2} \rangle$. ❑

We mention briefly that our proof approach clearly also yields $\Pi_2^p$-completeness for general graph families specified by the general-graph analog of tournament family specifiers, and for individual graphs specified in the Galperin–Wigderson formalism (see footnote 2). We now give the definitions and theorems to state this explicitly.

We say a function $f$ is a *graph family specifier* exactly if

1. $f$ is a polynomial-time computable function.

2. $(\forall x, y \in \Sigma^*)[f(x,y) = 1 \lor f(x,y) = 0]$.

We interpret this as specifying, in the following way, a family of (simple, directed) graphs. At each length $n$, the nodes in the graph specified by $f$ will be the strings in $\Sigma^n$. And for each two distinct nodes among these, $x$ and $y$, there is an edge from $x$ to $y$ exactly if $f(x,y) = 1$. We will call the graph just described *the length $n$ graph induced by $f$*.

We also define two sets to capture the complexity of kingship in tournaments and graphs in the Galperin–Wigderson model. The second of these sets, similarly to [Tan01, NT02], places into the complexity of the set the check (of coNP-type complexity, and easily handled) that the input circuit indeed is a tournament. Here, "the graph specified by $c$" of course refers to the Galperin–Wigderson model.

$$\text{Kings}_{GW} = \{\langle c, x \rangle \mid c \text{ has } 2|x| \text{ inputs and } x \text{ is a king in the graph specified by } c\}.$$

Tournament-Kings$_{GW}$ = $\{\langle c, x\rangle \,|\, c$ has $2|x|$ inputs and $x$ is a king in the graph specified by $c$ and the graph specified by $c$ is a tournament$\}$.

**Theorem 2.3**  1. There is a graph family specifier $f$ such that Kings$_f$ is $\Pi_2^p$-complete.

2. Kings$_{GW}$ is $\Pi_2^p$-complete.

3. Tournament-Kings$_{GW}$ is $\Pi_2^p$-complete.

Each of these problems is obviously in $\Pi_2^p$, and in light of our embedding of formulas into a king problem, is easily seen to be (though in parts 2 and 3 of Theorem 2.3 one has to be slightly careful about ensuring that we maintain a power of two cardinality of nodes) $\Pi_2^p$-complete.

Recall that each tournament family specifier $f$ satisfies Kings$_f \in \Pi_2^p$ [HOZZ], and Theorem 2.1 proves that for some tournament family specifier $\widehat{f}$, Kings$_{\widehat{f}}$ is $\Pi_2^p$-complete. What if we require our tournament-specifying functions to be not merely commutative but also associative? Does $\Pi_2^p$-completeness still hold? The answer is no, unless P = NP (see Theorem 2.5). Indeed, the complexity drops to coNP. However, beyond that, it also holds that the complexity in this case can never be coNP-complete unless P = NP. Also, the coNP upper bound here yields the result that for associatively P-selective sets, the $\Pi_2^p/1$ nonimmunity bound of [HT05] can be improved to coNP/1 nonimmunity. We now state and briefly prove these claims.

We say that a function $f$ is an *associative tournament family specifier* if it is a tournament family specifier and in addition satisfies

$$(\forall x, y, z \in \Sigma^*)[f(x, f(y, z)) = f(f(x, y), z)].$$

(In the study of P-selective sets, a set $A$ is said to be *associatively, commutatively P-selective* [HHN04] exactly if it has a P-selector function that is an associative tournament family selector $f$ and that in addition obeys the key constraint of P-selectivity: $(\forall x, y)[\{x, y\} \cap A \neq \emptyset \implies f(x, y) \in A]$. And a set is said to be *associatively P-selective* if it has a P-selector function that satisfies the same constraints, except without commutativity being required.)

**Theorem 2.4** *If $f$ is an associative tournament family specifier, then*

1. Kings$_f$ is a sparse set.

2. Kings$_f \in$ coNP.

**Proof**  Associativity of the specifier implies transitivity of each tournament in the family (i.e., in each, the edge set is transitive), and transitive tournaments impose a linear order on their nodes ([Moo68], see also [HHN04]). So each tournament in the family has exactly one king. So the king set is sparse; indeed, it has exactly one element per length. And due to the linear ordering, there is a coNP test for kingship in this case, namely, $z$ is a king exact if $(\forall y : |y| = |z|)[f(y, z) = z]$. ❑

**Theorem 2.5** *The following statements are equivalent:*

1. P = NP.
2. *Every associative tournament family specifier's king problem is* coNP*-complete.*
3. *Some associative tournament family specifier's king problem is* coNP*-complete.*
4. *Every associative tournament family specifier's king problem is* $\Pi_2^p$*-complete.*
5. *Some associative tournament family specifier's king problem is* $\Pi_2^p$*-complete.*

**Proof** The second part immediately implies the third part. The third part implies the first part by the sparseness claim of Theorem 2.4 plus the result of Fortune [For79] that if there exists a coNP-hard sparse set then P = NP. The first part implies the second part since if P = NP then every set other than $\emptyset$ and $\Sigma^*$ is coNP-complete (as always in this paper, with respect to $\leq_m^p$ reductions). But a king problem of an associative tournament family specifier, as noted in the proof of Theorem 2.4, always has exactly one king per length, and so the king set is neither $\emptyset$ nor $\Sigma^*$. The final two parts are equivalent to the first part by essentially the same reasoning, keeping in mind that all $\Pi_2^p$-hard sets are coNP-hard, and so Fortune's Theorem still applies. ❑

**Corollary 2.6 (Corollary to Theorem 2.4)** *If $L$ is an associatively* P*-selective, infinite set, then $L$ is not* coNP/1*-immune (i.e., $L$ has an infinite* coNP/1 *subset).*

**Proof** If $L$ is associatively P-selective, then by [HHN04] it is commutatively, associatively P-selective, say via function $f$. So by Theorem 2.4 its king set $\text{Kings}_f$ is in coNP. So infinite set $L$ has an infinite coNP/1 subset, namely, the coNP set is

$$\{\langle x, b\rangle \mid b = 1 \text{ and } x \text{ is the king in the length } |x| \text{ tournament induced by } f\},$$

the advice function at length $k$ outputs 1 if $\Sigma^k \cap L \neq \emptyset$ and outputs 0 if $\Sigma^k \cap L = \emptyset$, and the coNP/1 set thus is $\{y \mid y \in \text{Kings}_f \text{ and } \Sigma^{|y|} \cap L \neq \emptyset\}$. ❑

So, for some tournament family specifier $f$, $\text{Kings}_f$ is $\Pi_2^p$-hard, and for all associative tournament family specifiers $f$, $\text{Kings}_f \in$ coNP. But unless P = NP no associative tournament family specifier's king problem is coNP-complete. Do *any* tournament family specifiers have king problems that are coNP-complete? NP-complete? To give a sense of the possibilities that can hold, we prove that both of these cases hold. In doing so, we will exploit a property of the second part of our main construction—a property that was not critical there but that is critical here. In particular, the second part of our main construction—the part that weaves together an exponential number of subtournaments—does so in such a way as to not increase the complexity of king recognition for nodes in the subproblems: Each node associated with a subtournament is a king in the combined tournament exactly if it was a king in its native subtournament. Also useful will be that our $\Pi_2^p$ construction's new "bookkeeping" nodes that are employed in the "weaving together" stage of that construction have king problems that can be solved in coNP. The reason that these facts—that our weaving does not boost complexity—are important is that to prove, for example, coNP-completeness of $\text{Kings}_f$, we need not just coNP-hardness, but we need the king problem of $f$ to be *in* coNP. So, if for some nodes the king problem would require complexity beyond coNP, that would be fatal to our proof. The reason this was not an issue in our $\Pi_2^p$-completeness proof's construction is that $\Pi_2^p$ is an outright upper bound on king checking, and so that construction cannot possibly have nodes whose checking problem is beyond $\Pi_2^p$.

Since some of the bookkeeping nodes in our $\Pi_2^p$ construction's weaving stage seem to need coNP for their king tests, and that would ruin an NP upper bound, for our NP-completeness proof (Theorem 2.10) we will have to somewhat modify the weaving construction from that used in the $\Pi_2^p$ construction's weaving.

The issue of retaining the low upper bound of NP or coNP applies not just in the combination stage but also in the new—and aimed at coNP or NP—subtournaments we must design for the subtournament stages for those cases. In particular, when doing this research, the first subtournament we constructed for the NP case indeed had one node whose king problem supported NP-hardness and was in NP, but it also had other nodes whose king tests seemed to require the power of coNP. And thus we had to modify our subtournaments (from that first attempt) to add two extra nodes whose purpose was to make sure that all nodes have kingship problems that fall in NP.

**Theorem 2.7** *There exists a tournament family specifier $f$ such that $\text{Kings}_f$ is coNP-complete.*

**Proof** We will define a tournament family specifier $f$ such that $\text{Kings}_f$ is coNP-complete. For this tournament family specifier $f$, we will show coNP-hardness for $\text{Kings}_f$ by a reduction from TAUTOLOGY, the set of Boolean formulas that are tautologies, i.e., that are true for every assignment. Without loss of generality, we will in this proof assume that every formula contains at least one variable. We will denote the number of variables of formula $\phi$ by $n_\phi$.

As in the proof of Theorem 2.1, we will define our tournament specifier $f$ in two stages. First we will for every (propositional) formula $\phi$ with $n > 0$ variables define a tournament $T_\phi = (V_\phi, E_\phi)$ in such a way that $\langle \phi, 0^{n+2} \rangle$ is a king in $T_\phi$ if and only if $\phi$ is a tautology. (We use the same binary pairing function $\langle \cdot, \cdot \rangle$ as in the proof of Theorem 2.1.) We will then combine these tournaments into one family of tournaments (specified by $f$) using the construction from Theorem 2.1, and we will show that $\text{Kings}_f$ is coNP-complete. We use $\widehat{F}$ to denote all propositional formulas having at least one variable.

Let $\phi \in \widehat{F}$ be a formula with $n$ variables. Figure 3 gives a pictorial representation of $T_\phi$. The nodes in $V_\phi$ are arranged in three layers, and we use the convention that missing edges between nodes at different levels go "up," and that missing edges between nodes at the same level go "right."

Formally, $T_\phi = (V_\phi, E_\phi)$ is defined as follows. $V_\phi = \{\langle \phi, 0^{n+2} \rangle, \langle \phi, 100^n \rangle\} \cup \{\langle \phi, 11x \rangle \mid x \in \Sigma^n\}$. Note that all strings in $V_\phi$ have the same length by the properties of the pairing function. For all $z, z' \in \Sigma^{n+2}$ such that $\langle \phi, z \rangle, \langle \phi, z' \rangle \in V_\phi$ and $z < z'$, let $(\langle \phi, z \rangle, \langle \phi, z' \rangle) \in$

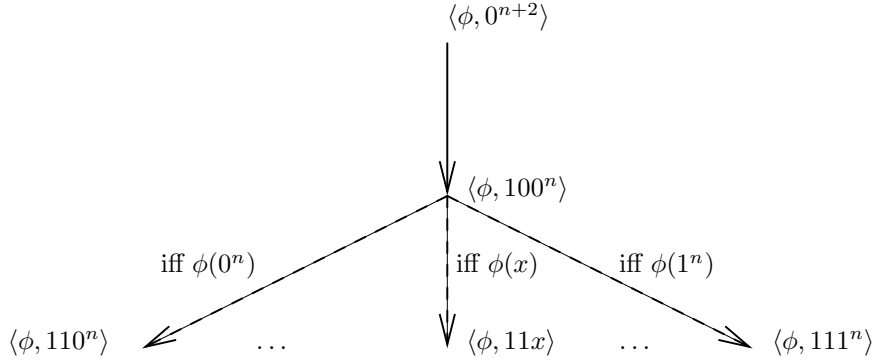

Figure 3: First stage of the coNP-completeness construction

$E_\phi$ if and only if

- ($z = 0^{n+2}$ and $z' = 100^n$), or
- ($z = 100^n$ and $z' = 11x'$ for some $x' \in \Sigma^n$ and $\phi(x')$), or
- ($z = 11x$ for some $x \in \Sigma^n$ and $z' = 11x'$ for some $x' \in \Sigma^n$).

For all $z, z' \in \Sigma^{n+2}$ such that $\langle\phi, z\rangle, \langle\phi, z'\rangle \in V_\phi$ and $z > z'$, let $(\langle\phi, z\rangle, \langle\phi, z'\rangle) \in E_\phi$ if and only if $(\langle\phi, z'\rangle, \langle\phi, z\rangle) \notin E_\phi$.

It is clear that $T_\phi$ is a tournament. The following claim follows immediately from the construction.

**Claim 2.8** $\langle\phi, 0^{n+2}\rangle$ *is a king in tournament $T_\phi$ if and only if $\phi$ is tautology.*

It is also immediate that in tournament $T_\phi$,

- $\langle\phi, 100^n\rangle$ is a king if and only if $\phi(0^n)$ (if $\neg\phi(0^n)$, then $\langle\phi, 110^n\rangle$ is not reachable from $\langle\phi, 100^n\rangle$) and
- for all $x \in \Sigma^n$, $\langle\phi, 11x\rangle$ is a king if and only if $(\neg\phi(x)) \wedge (\forall x' \in \Sigma^n)[x' < x \Rightarrow \phi(x')]$.

From these observations and the polynomial-time invertibility of the pairing function, we have the following.

**Claim 2.9** $\{z \mid (\exists \phi \in \widehat{F})[z \text{ is a king in tournament } T_\phi]\}$ *is in* coNP.

To combine the $T_\phi$ tournaments into a family of tournaments specified by $f$, we use the same construction as that in the proof of Theorem 2.1 (as depicted in Figure 2), except that we replace in the construction every occurrence of the string "∀∃-formula" by the string "formula in $\widehat{F}$."

The same argument as in the proof of Theorem 2.1 can be used to show that $f$ is a tournament family specifier and that for every formula $\phi \in \widehat{F}$, $\langle\phi, 0^{n_\phi+2}\rangle \in \text{Kings}_f$ if and only if $\phi$ is a tautology. This shows that $\text{Kings}_f$ is coNP-hard.

It remains to show that $\text{Kings}_f$ is in coNP. The following hold:

- For all $m$, $0^m \in \text{Kings}_f$.

- For all $z \in \textit{Other}$, $z \notin \text{Kings}_f$ (since $0^{|z|}$ is not reachable from $z$).

- For all $z$ such that $(\exists \phi \in \widehat{F})[z \in V_\phi]$, $z \in \text{Kings}_f$ if and only if $z$ is a king in $T_\phi$, since there are no paths of the form $v \to w \to v'$ such that $v, v' \in V_\phi$ and $w \notin V_\phi$.

- For all $z$ such that $(\exists \phi \in \widehat{F})[z = \langle \phi, 010^{n_\phi} \rangle]$, $z \in \text{Kings}_f$ if and only if $\phi$ is the lexicographically smallest formula such that $|\langle \phi, 0^{n_\phi+2} \rangle| = m$, i.e., if and only if for all formulas $\psi < \phi$, it holds that $|\langle \psi, 0^{n_\psi+2} \rangle| > m$.

From these observations and Claim 2.9 it follows immediately that $\text{Kings}_f$ is in coNP. ❑

**Theorem 2.10** *There exists a tournament family specifier $f$ such that $\text{Kings}_f$ is NP-complete.*

**Proof** We will define a tournament family specifier $f$ such that $\text{Kings}_f$ is NP-complete. For this tournament family specifier $f$, we will show NP-hardness for $\text{Kings}_f$ by a reduction from SAT, the set of satisfiable Boolean formulas. As in the proof of Theorem 2.7, we will in this proof without loss of generality assume that every formula contains at least one variable. We will denote the number of variables of formula $\phi$ by $n_\phi$.

As in the proofs of Theorems 2.1 and Theorem 2.7, we will define our tournament specifier $f$ in two stages. First we will for every (propositional) formula $\phi$ with $n > 0$ variables define a tournament $T_\phi = (V_\phi, E_\phi)$ in such a way that $\langle \phi, 0^{n+2} \rangle$ is a king in $T_\phi$ if and only if $\phi$ is satisfiable. We will then show how to combine these tournaments into one family of tournaments (specified by $f$) such that $\text{Kings}_f$ is NP-complete. In contrast to the proof of Theorem 2.7, we cannot use the exact same construction as that of Theorem 2.1, since in that construction, $\langle \phi, 010^{n_\phi} \rangle \in \text{Kings}_f$ if and only if $\phi$ is the lexicographically smallest formula such that $|\langle \phi, 0^{n_\phi+2} \rangle| = m$, i.e., if and only if for all formulas $\psi < \phi$, $|\langle \psi, 0^{n_\psi+2} \rangle| > m$. This is a coNP predicate, but we need the king problem to be in NP. We will show in the sequel how to modify the construction from Theorem 2.1. In order to be able to do so, we modify the pairing function from the proof of Theorem 2.1 such that it obeys the extra requirement that the pairing function never outputs a string in $0^*1 + 10^*$. It is easy to see that such a pairing function exists, for example by defining $\langle x_1 x_2 \cdots x_n, y \rangle$ as $0x_1 0x_2 \cdots 0x_n 11y$ for all $x_1, x_2, \ldots, x_n \in \Sigma$ and $y \in \Sigma^*$.

Let $\phi \in \widehat{F}$ be a formula with $n$ variables. $\widehat{F}$ again denotes the set of all propositional formulas having at least one variable. Figure 4 gives a pictorial representation of $T_\phi$. The nodes in $V_\phi$ are arranged in five layers, and we use the convention that missing edges between nodes at different levels go "up," and that missing edges between nodes at the same level go "right." Note that since $n > 0$, the layers are disjoint.

Formally, $T_\phi = (V_\phi, E_\phi)$ is defined as follows. $V_\phi = \{\langle \phi, 0^{n+2} \rangle, \langle \phi, 110^n \rangle, \langle \phi, 001^n \rangle, \langle \phi, 1^{n+2} \rangle\} \cup \{\langle \phi, 10x \rangle \mid x \in \Sigma^n\}$. Note that all strings

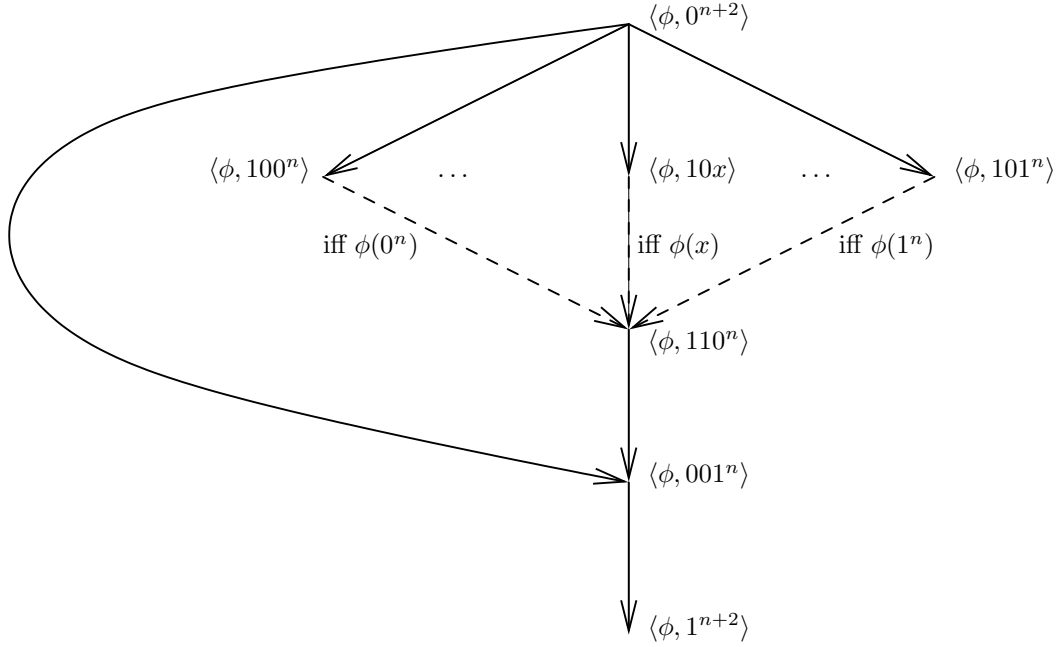

Figure 4: First stage of the NP-completeness construction

in $V_\phi$ have the same length by the properties of the pairing function. For all $z, z' \in \Sigma^{n+2}$ such that $\langle \phi, z \rangle, \langle \phi, z' \rangle \in V_\phi$ and $z < z'$, let $(\langle \phi, z \rangle, \langle \phi, z' \rangle) \in E_\phi$ if and only if

- ($z = 0^{n+2}$ and $z' = 001^n$), or
- ($z = 0^{n+2}$ and $z' = 10x$ for some $x \in \Sigma^n$), or
- ($z = 001^n$ and $z' = 10x$ for some $x \in \Sigma^n$), or
- ($z = 001^n$ and $z' = 1^{n+2}$), or
- ($z = 10x$ for some $x \in \Sigma^n$ and $z' = 10x'$ for some $x \in \Sigma^n$), or
- ($z = 10x$ for some $x \in \Sigma^n$ and $z' = 110^n$ and $\phi(x)$).

For all $z, z' \in \Sigma^{n+2}$ such that $\langle \phi, z \rangle, \langle \phi, z' \rangle \in V_\phi$ and $z > z'$, let $(\langle \phi, z \rangle, \langle \phi, z' \rangle) \in E_\phi$ if and only if $(\langle \phi, z' \rangle, \langle \phi, z \rangle) \notin E_\phi$.

It is clear that $T_\phi$ is a tournament. It is immediate from the construction that the following holds.

**Claim 2.11** $\langle \phi, 0^{n+2} \rangle$ *is a king in tournament $T_\phi$ if and only if $\phi$ is satisfiable.*

Note that Claim 2.11 also holds in the simpler tournament that results when we remove nodes $\langle \phi, 001^n \rangle$ and $\langle \phi, 1^{n+2} \rangle$ from $T_\phi$. However, in this simpler tournament, $\langle \phi, 101^n \rangle$ is a king if and only if for all $x \in \Sigma^n$, $\phi(x) \Leftrightarrow x = 1^n$. This is a coNP predicate, but we need the king problem to be in NP.

In $T_\phi$, the following hold.

- $\langle\phi, 001^n\rangle$, $\langle\phi, 110^n\rangle$, and $\langle\phi, 1^{n+2}\rangle$ are kings.
- For all $x \in \Sigma^n$, $\langle\phi, 10x\rangle$ is not a king, since $\langle\phi, 1^{n+2}\rangle$ is not reachable from $\langle\phi, 10x\rangle$ by a path of length at most two.

From these observations, Claim 2.11, and the polynomial-time invertibility of the pairing function, we have the following claim.

**Claim 2.12** $\{z \mid (\exists \phi \in \widehat{F})[z \text{ is a king in tournament } T_\phi]\}$ *is in* NP.

As explained earlier in this proof, we cannot use the construction from the proof of Theorem 2.1 to combine the $T_\phi$ tournaments into a family of tournaments with the desired properties. We will modify the construction from the proof of Theorem 2.1 as follows. For all $m \geq 2$,

- Remove two nodes $0^{m-1}1$ and $10^{m-1}$ from the set *Other*. (Recall that we modified the pairing function in such a way that these two strings are not in the range of the pairing function.)
- For all $z \in \Sigma^m$, let $f(0^{m-1}1, z) = 0^{m-1}1$ if and only if $z \in \{0^{m-1}1, 10^{m-1}\} \cup \textit{Other} \cup \{\langle\phi, 010^{n_\phi}\rangle \mid \phi \in \widehat{F}\}$.
  For all $z \in \Sigma^m$, if $f(0^{m-1}1, z) \neq 0^{m-1}1$, then let $f(0^{m-1}1, z) = z$.
- For all $z \in \Sigma^m$, let $f(10^{m-1}, z) = 10^{m-1}$ if and only if $z \neq 0^{m-1}1$.
- The rest of the construction remains the same, except that, as in the proof of Theorem 2.7, we replace every occurrence of the string "$\forall\exists$-formula" by the string "formula in $\widehat{F}$."

Arguments similar to those in the proof of Theorem 2.1 clearly show that $f$ is a tournament family specifier and that for every formula $\phi \in \widehat{F}$, $\langle\phi, 0^{n_\phi+2}\rangle \in \text{Kings}_f$ if and only if $\phi$ is satisfiable. This shows that $\text{Kings}_f$ is NP-hard.

It remains to show that $\text{Kings}_f$ is in NP. The following hold.

- For all $m$, $0^m \in \text{Kings}_f$.
- For all $m \geq 2$, $0^{m-1}1 \in \text{Kings}_f$, and $10^{m-1} \in \text{Kings}_f$.
- For all $z$ such that $|z| \geq 2$ and $z \in \textit{Other}$, $z \notin \text{Kings}_f$ (since $0^{|z|}$ is not reachable from $z$).
- For all $z$ such that $|z| \geq 2$ and $(\exists\phi \in \widehat{F})[z \in V_\phi]$, $z \in \text{Kings}_f$ if and only if $z$ is a king in $T_\phi$ (since there are no paths of the form $v \to w \to v'$ such that $v, v' \in V_\phi$ and $w \notin V_\phi$).
- For all $z$ such that $|z| \geq 2$ and $(\exists\phi \in \widehat{F})[z = \langle\phi, 010^{n_\phi}\rangle]$, it holds that $z \notin \text{Kings}_f$ (since $10^{m-1}$ is not reachable from $z$ by a path of length at most two).

From these observations and Claim 2.12 it follows immediately that $\text{Kings}_f$ is in NP. ❑

## 3 The Complexity of $k$-Kingship in Tournaments and Graphs

The previous section studied the complexity of kingship (i.e., 2-kingship) in tournaments and graphs. In this section we study, in tournaments and graphs, the complexity of $k$-kingship for $k = 1$ and for $k > 2$. The highlight of this section is Theorem 3.4, which shows that for $k > 2$, $k$-kingship remains $\Pi_2^p$-complete. To show $\Pi_2^p$-completeness, our proof draws on the 2-king subtournament part of the construction from Theorem 2.1's proof, plus a new combination phase that involves adding antennas and performing appropriate interweaving.

We will refer to the following sets. For each positive integer $k$ and each tournament family specifier $f$, define

$$k\text{-Kings}_f = \{x \mid x \text{ is a } k\text{-king in the length } |x| \text{ tournament induced by } f\}.$$

For each positive integer $k$ define the following two sets.

$k\text{-Kings}_{GW} = \{\langle c, x \rangle \mid c \text{ has } 2|x| \text{ inputs and } x \text{ is a } k\text{-king in the graph specified by } c\}$.

$k\text{-Tournament-Kings}_{GW} = \{\langle c, x \rangle \mid c \text{ has } 2|x| \text{ inputs and } x \text{ is a } k\text{-king in the graph specified by } c \text{ and the graph specified by } c \text{ is a tournament}\}$.

### 3.1 1-Kingship in Tournaments and Graphs

Let us first quickly discuss the quite simple case of 1-kings. Note that in a tournament, a node is a 1-king exactly if it points to each other node. This is easy to test in coNP. The "points to each other node" observation makes it clear that a given tournament has a most one 1-king. Thus, if we study 1-kingship in the tournament family specifier model, we have the following result.

**Theorem 3.1**  1. *For each tournament family specifier $f$, $1\text{-Kings}_f \in \text{coNP}$.*

2. *Unless $\text{P} = \text{NP}$, for no tournament family specifier $f$ is $1\text{-Kings}_f$ coNP-complete.*

On the other hand, the natural analog of $\text{Kings}_{GW}$ for 1-kingship (both in tournaments and in general graphs) easily yields coNP-completeness. The upper bound of coNP is immediate, and we can easily build a circuit with a header node and "potential certificate nodes" and have the header node point to the certificates, and so can test tautology. (As always, we will need some dummy nodes, always pointed to by the header, to ensure that the whole graph's size is a power of two, but this is a minor detail. The cross-arrows between potential certificate nodes can be set arbitrarily, as can the cross-arrows between those and the padding nodes.) Thus, we have first part of the following result (and the second part is similarly easy to see).

**Theorem 3.2** *$1\text{-Kings}_{GW}$ and $1\text{-Tournament-Kings}_{GW}$ are coNP-complete.*

## 3.2 $k$-Kingship in Tournaments and Graphs, $k > 2$

In this section, we study the complexity of $k$-kingship, $k > 2$, in tournaments and graphs.

For clarity, we first quickly directly dispatch the case of individually specified tournaments and graphs (though we could argue it indirectly via family-specified tournaments). Then we will turn to the far more interesting case of tournaments specified via a tournament family specifier.

For both individually specified graphs and tournaments, $k$-kingship, $k \geq 2$, is $\Pi_2^p$-complete. This follows from an easy "antenna adding" modification to the construction used in the first stage of the proof of Theorem 2.1.

**Theorem 3.3** *For each $k \geq 2$, $k$-Kings$_{GW}$ and $k$-Tournament-Kings$_{GW}$ are $\Pi_2^p$-complete.*

**Proof** In both cases, the $\Pi_2^p$ upper bound is not hard to see. The $\Pi_2^p$ lower bound for $k$-Kings$_{GW}$ essentially follows from obtaining a $\Pi_2^p$ lower bound for $k$-Tournament-Kings$_{GW}$, so we address just that issue. Consider some fixed $k \geq 2$. As noted in the final two parts of the statement of Theorem 2.3, the first stage construction from the proof of Theorem 2.1 in effect shows the $k = 2$ case. If $k > 2$, let us describe the succinctly specified tournament that handles that case. The tournament will add a directed chain of $k - 2$ nodes, the last of which will point to the node-being-checked-for-kingship of the $k = 2$ construction. Except for the one arrow just mentioned, which connects the end of the chain to the original part, all nodes from the original part of the construction point to all the nodes of this chain. And aside from the forward-pointing arrows in the chain itself, all other arrows between nodes in the chain point from the node nearer the original part to the node nearer the chain's first node. Note that the first node of the chain clearly will be a $k$-king exactly if the node-being-checked-for-kingship of the $k = 2$ construction is a 2-king. (Though we said above we would leave tacit issues of padding nodes, let us mention them explicitly here as an example. The number of nodes in the construction just given in general might not be a power of two. In that case, we pad up to the next power of two with dummy nodes, all of which are pointed to by all the nodes in our construction, and that point amongst themselves in any fixed, easy-to-compute way.) ❏

We now come to the main result of this section, the complexity of $k$-Kings$_f$, $k > 2$. Recall the two-stage approach we employed in Section 2: In the first stage of our construction, we showed how a 2-king problem could capture a single $\Pi_2^p$-type formula. In the second stage, we showed how to very uniformly combine an exponential number of such subtournaments into a giant tournament in which each potential 2-king in each subtournament has the property that it is a 2-king in its subtournament if and only if it is a 2-king in the new combined tournament.

It would be natural to hope that the same attack would work for 3-kings, 4-kings, and so on. However, there are serious obstacles to that approach. By a somewhat tricky argument, using in part the nice fact that in a tournament each node that is on a long cycle in fact is on some 3-cycle, one can see that if one substitutes (for example) "3" for "2" in the above plan of attack, and one makes a few other assumptions about uniformity of interrelations between subtournaments (an assumption about cyclic connections that is satisfied by our 2-kings construction, and most critically, that given any two nodes $a$ and $b$ from a subtournament and any node $c$ not in that subtournament, either both $a$ and $b$ point to $c$ or both $a$ and $b$ are pointed to by $c$; this is a strong assumption but note that our 2-kings combined tournament construction in fact satisfies it), then every such construction in fact makes each subtournament's potential-3-king always be a 3-king in its subtournament within the combined tournament (via a cross-pollution path). So, this at least hints that we may not be able to just clone the 2-stage construction approach, and first show that 3-kings can handle one $\Pi_2^p$-type problem, and then show how to combine an exponential number of 3-king examples uniformly.

Rather, we do the following. For our subtournaments, we use not 3-king problems but 2-king problems. After all, we already know that even 2-king problems can encode $\Pi_2^p$-type formulas! And then we in our combination stage combine things so as to ensure that each former potential-2-king has a related node that will be a 3-king in the overall tournament exactly if the potential-2-king is a 2-king in the original subtournament. Very informally put, instead of using a 3-king specific first stage, we steal that from the 2-king construction. And our second stage sticks appropriate length antennas onto each subtournament and appropriately weaves them all together. (Regarding the obstacle mentioned above, one might ask why the obstacle does not apply if one views each 2-king subtournament and its antenna as a 3-king subtournament. The answer is that our weaving treats the body of the subtournament and the subtournament's antenna differently in how they connect to other nodes, and that violates the $a/b/c$ uniformity mentioned above. So, our construction thus bypasses that obstacle.)

We now state and prove this result, which generalizes Theorem 2.1, but which also draws in part on the construction used to prove Theorem 2.1.

**Theorem 3.4** *For each $k \geq 2$, there exists a tournament family specifier $f$ such that $k$-Kings$_f$ is $\Pi_2^p$-complete.*

**Proof** Consider some fixed $k \geq 2$. It is easy to see that for every tournament family specifier $f$, $k$-Kings$_f$ is in $\Pi_2^p$. We will construct a tournament family specifier $f$ such that $k$-Kings$_f$ is $\Pi_2^p$-hard. Similarly to the proof of Theorem 2.1, we will show $\Pi_2^p$-hardness for $k$-Kings$_f$ by a reduction from a variant of $\forall \exists$SAT. All formulas in the variant of $\forall \exists$SAT will have the same number of universally quantified variables as existentially quantified variables and their number of universally quantified variables must be greater than $k - 2$. We will call formulas of the right form $\forall \exists_k$-formulas, i.e., we say that $\phi$ is a $\forall \exists_k$-formula if there exists an integer $n > k - 2$ and a propositional formula $\phi'$ such that $\phi = \forall x_1 \cdots \forall x_n \exists y_1 \cdots \exists y_n \phi'(x_1, \ldots, x_n, y_1, \ldots, y_n)$. $\forall \exists$SAT$_k$ will denote the set of true $\forall \exists_k$-formulas. For standard reasons, $\forall \exists$SAT$_k$ is $\Pi_2^p$-complete.

We now will define a tournament family specifier $f$ (dependent on $k$) and then will argue that $\forall \exists$SAT$_k \leq_m^p k$-Kings$_f$.

Let $\langle \cdot, \cdot \rangle$ be the pairing function from the proof of Theorem 2.1. For $\phi$ a $\forall \exists_k$-formula, let $T_\phi = (V_\phi, E_\phi)$ be the tournament of that name from the proof of Theorem 2.1. Recall (Claim 2.2) that for every $\forall \exists_2$-formula $\phi$, $\langle \phi, 0^{n_\phi + 2} \rangle$ is a 2-king in tournament $T_\phi$ if and only if $\phi$ is true. (As in the proof of Theorem 2.1, $n_\phi$ denotes the number of universally quantified variables in $\phi$.) So, it certainly also holds that for each $\forall\exists_k$-formula $\phi$, $\langle\phi, 0^{n_\phi+2}\rangle$ is a 2-king in tournament $T_\phi$ if and only if $\phi$ is true (since each $\forall\exists_k$-formula is a $\forall\exists_2$-formula).

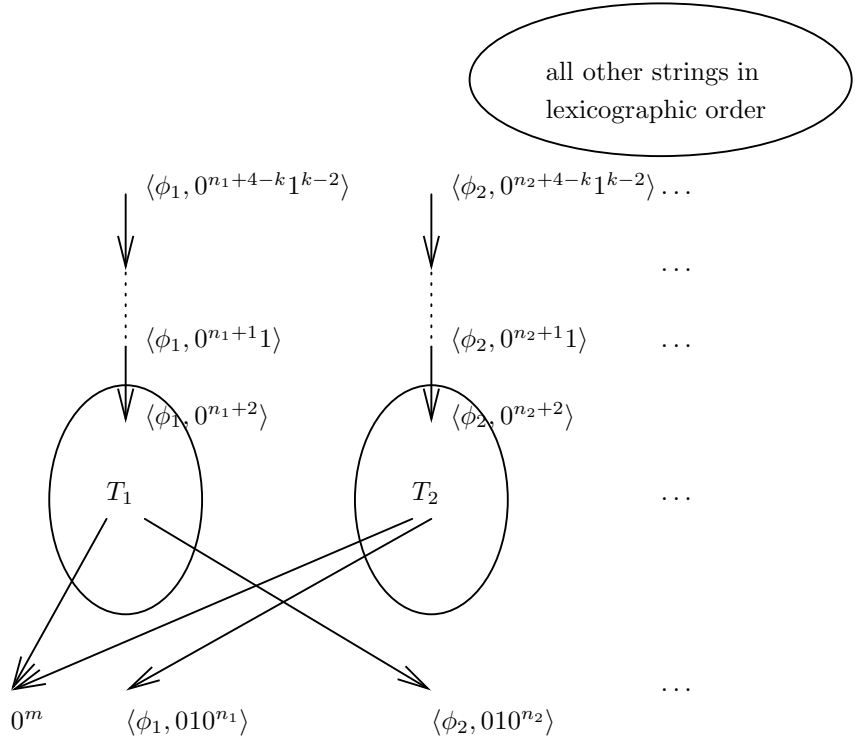

Figure 5: Second stage of the $k$-Kings$_f$ $\Pi_2^p$-hardness construction

We will combine the $T_\phi$ tournaments into a family of tournaments specified by $f$ in such a way that for every $\forall\exists_k$-formula $\phi$, $\langle\phi, 0^{n_\phi+2}\rangle$ is a 2-king in $T_\phi$ if only if $\langle\phi, 0^{n_\phi+4-k}1^{k-2}\rangle \in k$-Kings$_f$.

Figure 5 gives a pictorial representation of the tournament induced by $f$ at length $m$. In the figure, $\phi_1, \phi_2, \ldots, \phi_k$ are all $\forall\exists_k$-formulas such that $V_{\phi_i} \subseteq \Sigma^m$. The $\phi_i$'s are ordered lexicographically, in ascending order. For readability, we write $T_i$ for $T_{\phi_i}$ and $n_i$ for $n_{\phi_i}$. Note that for all formulas $\phi$, $0^m \notin V_\phi$ (by properties of the pairing function), $\langle\phi, 010^{n_\phi}\rangle \notin V_\phi$, and that for all $i$, $1 \leq i \leq k-2$, $\langle\phi, 0^{n+2-i}1^i\rangle \notin V_\phi$. We use the convention that all missing arrows between nodes at the same level of Figure 5 go right (we assume that all nodes in $T_i$ are at the same level, and that all nodes in "all other strings" are at the same level) and that all other missing arrows go up. This completely specifies the tournament on strings of length $m$. Note that for $k = 2$, this is the exact same tournament as that depicted in Figure 2.

Formally, we define $f$ as follows. Let

$$Antenna = \{\langle \phi, 0^{n_\phi+2-i}1^i \rangle \mid \phi \text{ is a } \forall\exists_k\text{-formula and } 1 \leq i \leq k-2\}$$

and let

$$Other = \Sigma^* - \left(0^* \cup \{\langle \phi, 010^{n_\phi}\rangle \mid \phi \text{ is a } \forall\exists_k\text{-formula}\} \cup Antenna \cup \bigcup \{V_\phi \mid \phi \text{ is a } \forall\exists_k\text{-formula}\}\right).$$

The sets *Antenna* and *Other* are clearly in P, since the pairing function is polynomial-time invertible. For all $m$ and all $z, z' \in \Sigma^m$, let $f(z, z') = z$ if and only if

- $z = z'$, or
- $z = 0^m$ and $z' = \langle \phi, 010^{n_\phi}\rangle$ for some $\forall\exists_k$-formula $\phi$, or
- $z = 0^m$ and $z' \in Antenna \cup Other$, or
- $z = \langle \phi, 010^{n_\phi}\rangle$ for some $\forall\exists_k$-formula $\phi$ and $z' = \langle \psi, 010^{n_\psi}\rangle$ for some $\forall\exists_k$-formula $\psi$ and $\phi < \psi$, or
- $z = \langle \phi, 010^{n_\phi}\rangle$ for some $\forall\exists_k$-formula $\phi$ and $z' \in V_\phi$, or
- $z = \langle \phi, 010^{n_\phi}\rangle$ for some $\forall\exists_k$-formula $\phi$ and $z' \in Antenna \cup Other$, or
- $z \in V_\phi$ for some $\forall\exists_k$-formula $\phi$ and ($z' = 0^m$ or $z' \in Other$), or
- $z \in V_\phi$ for some $\forall\exists_k$-formula $\phi$ and $z' = \langle \psi, 010^{n_\psi}\rangle$ for some $\forall\exists_k$-formula $\psi$ and $\phi \neq \psi$, or
- $z \in V_\phi$ for some $\forall\exists_k$-formula $\phi$ and $z' \in V_\phi$ and $(z, z') \in E_\phi$, or
- $z \in V_\phi$ for some $\forall\exists_k$-formula $\phi$ and $z' \in V_\psi$ for some $\forall\exists_k$-formula $\psi$ and $\phi < \psi$, or
- $z \in V_\phi - \{\langle \phi, 0^{n_\phi+2}\rangle\}$ for some $\forall\exists_k$-formula $\phi$ and $z' \in Antenna$, or
- $z = \langle \phi, 0^{n_\phi+2}\rangle$ for some $\forall\exists_k$-formula $\phi$ and $z' \in Antenna - \{\langle \phi, 0^{n_\phi+1}1\rangle\}$, or
- $z = \langle \phi, 0^{n_\phi+2-i}1^i\rangle$ for some $\forall\exists_k$-formula $\phi$ and some $i$ such that $1 \leq i \leq k-2$ and $z' = \langle \phi, 0^{n_\phi+3-i}1^{i-1}\rangle$, or
- $z = \langle \phi, 0^{n_\phi+2-i}1^i\rangle$ for some $\forall\exists_k$-formula $\phi$ and some $i$ such that $1 \leq i \leq k-2$ and $z' = \langle \psi, 0^{n_\psi+2-i}1^i\rangle$ for some $\forall\exists_k$-formula $\psi$ and $\phi < \psi$, or
- $z = \langle \phi, 0^{n_\phi+2-i}1^i\rangle$ for some $\forall\exists_k$-formula $\phi$ and some $i$ such that $1 \leq i \leq k-2$ and $z' = \langle \psi, 0^{n_\psi+2-j}1^j\rangle$ for some $\forall\exists_k$-formula $\psi$ and some $j$ such that $1 \leq j \leq k-2$ and $i < j$, or
- $z = \langle \phi, 0^{n_\phi+2-i}1^i\rangle$ for some $\forall\exists_k$-formula $\phi$ and some $i$ such that $1 \leq i \leq k-2$ and $z' \in Other$, or
- $z, z' \in Other$ and $z < z'$.

For all $z, z' \in \Sigma^m$, if $f(z, z') \neq z$, then let $f(z, z') = z'$. The definition of $f$ on strings of different lengths is irrelevant (as long as $f$ remains a tournament family specifier). To be complete, we define $f(z, z') = z$ if $|z| < |z'|$ and $f(z, z') = z'$ if $|z| > |z'|$. This completes the definition of $f$.

It is immediate that for all $x, y \in \Sigma^*$, $f(x, y) = f(y, x)$ and that $f(x, y) = x$ or $f(x, y) = y$. Since the pairing function is polynomial-time invertible, it is easy to see that $f$ is computable in polynomial time. Thus $f$ is indeed a tournament family specifier.

We now define a $\leq_m^p$-reduction $g$ via which $\forall \exists \text{SAT}_k \leq_m^p k\text{-Kings}_f$. Note that $Other \neq \emptyset$ and no element of $Other$ belongs to $k\text{-Kings}_f$. Let $out$ be any fixed element of $Other$. Let $\phi$ be any element of $\Sigma^*$. Our reduction $g$ on input $\phi$ will output $out$ if $\phi$ is not a $\forall \exists_k$-formula. Otherwise, letting $n$ denote $n_\phi$, our reduction $g$ will output $\langle \phi, 0^{n+4-k}1^{k-2}\rangle$. In light of the fact, mentioned earlier, that $\langle \phi, 0^{n_\phi+2}\rangle$ is a 2-king in tournament $T_\phi$ if and only if $\phi$ is true, all that remains to show is that $\langle \phi, 0^{n+4-k}1^{k-2}\rangle \in k\text{-Kings}_f$ if and only if $\langle \phi, 0^{n+2}\rangle$ is a 2-king in $T_\phi$. Let $m = |\langle \phi, 0^{n+2}\rangle|$.

First suppose that $\langle \phi, 0^{n+2}\rangle$ is a 2-king in $T_\phi$. Then $\langle \phi, 0^{n+2}\rangle$ reaches all strings in $V_\phi$ in at most two steps. Also, as in the proof of Theorem 2.1, it is easy to see from the definition of $f$ that $\langle \phi, 0^{n+2}\rangle$ reaches all strings in $\Sigma^m - V_\phi$ in one or two steps. Since there is a path of length $k - 2$ from $\langle \phi, 0^{n+4-k}1^{k-2}\rangle$ to $\langle \phi, 0^{n+2}\rangle$, it follows that every string of length $m$ can be reached from $\langle \phi, 0^{n+4-k}1^{k-2}\rangle$ by a path of length at most $k$. It follows that $\langle \phi, 0^{n+4-k}1^{k-2}\rangle \in k\text{-Kings}_f$.

For the converse, suppose that $\langle \phi, 0^{n+4-k}1^{k-2}\rangle$ is a $k$-king in the tournament induced by $f$ on strings of length $m$. Then every string of length $m$ can be reached from $\langle \phi, 0^{n+4-k}1^{k-2}\rangle$ in at most $k$ steps. Fix a $v \in V_\phi$. There exists a $v' \in \Sigma^m$ such that $v$ can be reached in at most two steps from $v'$ and $v'$ can be reached in at most $k - 2$ steps from $\langle \phi, 0^{n+4-k}1^{k-2}\rangle$. Consider the possibilities for $v'$. From the definition of $f$, there are only four kinds of nodes reachable from $\langle \phi, 0^{n+4-k}1^{k-2}\rangle$ in at most $k - 2$ steps.

- $v' \in Other$. In this case, $v$ is not reachable from $v'$.

- $v' = \langle \phi, 0^{n+2-i}1^i\rangle$ for some $i$ such that $0 \leq i \leq k - 2$. In this case, it is clear that since $v$ is reachable from $v'$ in at most two steps, $v$ is reachable from $\langle \phi, 0^{n+2}\rangle$ in at most two steps.

- $v' = \langle \psi, 0^{n_\psi+2-i}1^i\rangle$ for some $\forall \exists_k$-formula $\psi > \phi$ and for some $i$ such that $1 \leq i \leq k-2$. In this case, $v$ is not reachable from $v'$ by a path of length at most 2.

- $v' = \langle \psi, 0^{n_\psi+2-i}1^i\rangle$ for some $\forall \exists_k$-formula $\psi < \phi$ and for some $i$ such that $2 \leq i \leq k-2$. In this case, $v$ is not reachable from $v'$ by a path of length at most 2.

From this case distinction, it follows that $v$ is reachable by a path of length at most two from $\langle \phi, 0^{n+2}\rangle$. Since there do not exist paths $\langle \phi, 0^{n+2}\rangle \to w \to v$ such that $w \notin V_\phi$ in the tournament induced by $f$ on length $m$, it follows $v$ can be reached from $\langle \phi, 0^{n+2}\rangle$ by a path of length at most two such that all nodes on the path are in $V_\phi$. It follows that $\langle \phi, 0^{n+2}\rangle$ is a 2-king in the tournament induced by $f$ on $V_\phi$, and thus $\langle \phi, 0^{n+2}\rangle$ is a 2-king in $T_\phi$. ❑

It is not hard to see that, as a corollary to the proof's method, one can even establish the following slightly stronger claim: There exists a tournament family specifier $f$ such that for all $k \geq 2$, $k$-Kings$_f$ is $\Pi_2^p$-complete. (Briefly put, one does this by at longer and longer lengths using longer and longer antennas and, for lengths at which the antennas are long enough to handle a particular $k$, mapping to the appropriate node within the appropriate antenna, and by, for the finite number of lengths that don't have long enough antennas to handle that $k$, using table lookup.)

# 4 The Complexity of $k$-Kingship in Multipartite Tournaments

In this section, we discuss the complexity of $k$-kingship in $j$-partite tournaments. (For each $j > 1$, a $j$-partite tournament is a completely oriented $j$-partite simple digraph. That is, between each pair of nodes in the same one of the $j$ parts, there are no edges. And if $v$ and $w$ are in different ones of the $j$ parts, then exactly one of the edges $(v, w)$ and $(w, v)$ belongs to the edge set.)

The first issue to address is one of model. Recall that Section 2 focused on tournament *family* specifiers. The reason it did so is because the motivation for that study was the theory of P-selective sets, where the key tool is precisely such families of tournament generators. However, as we study *multipartite* tournaments (which are where 3-kingship, 4-kingship and so on have mostly been discussed in the literature), that connection no longer exists. And so we will focus not on family specifiers but rather will focus on taking graphs, specified succinctly as circuits, as our input—or, to be more specific, taking as our input such a graph plus the node whose $k$-kingship we wish to investigate.

The multipartite setting raises another issue regarding model. The issue is: Should the burden of ensuring that the circuit indeed is computing a legal multipartite tournament be wedged into the complexity of the set we define, or should our model of specifying multipartite tournaments via circuits be such as to inherently ensure that we always define legal multipartite tournaments? Both approaches arguably have merits. However, since we wish to focus our attention on the complexity of kingship decisions, we make the latter choice: Our model of specifying multipartite tournaments via circuits will be such as to inherently ensure that we always define legal multipartite tournaments. This is also in line with the choice—made in the sets already defined in this paper regarding tournaments (except the sets Tournament-Kings$_{GW}$ and $k$-Tournament-Kings$_{GW}$, which take the former approach)—to inherently ensure that our graphs are tournaments. At the end of this section, we make some comments on the other approach.

Finally, in the previous sections we were rather rigorous in our constructions. We generally included not just pictures, but in our text specified exactly what nodes would be encoded as what strings, and exactly how padding/garbage nodes (such as those needed to make the total number of nodes come to a power of two) would be handled. In this section, in cases where it will not lead to confusion, for conciseness and clarity as to our key ideas, we sometimes are somewhat more informal in the way we state our constructions, and we sometimes leave out minor details such as padding nodes in cases when it is not hard to see how to easily handle them.

So we now turn to our results on $k$-kingship in multipartite ($j$-partite) tournaments. As mentioned above, we formalize our model so as to have the model itself ensure that every circuit we consider indeed defines a $j$-partite tournament.

In particular, we will say that a circuit is a *$j$-tournament-circuit* if for some $n$ it has exactly $j(n+1)$ inputs, and of course one output wire. We interpret this as specifying a $j$-partite tournament (a self-loop-free, complete, oriented $j$-partite graph) as follows. Our $j$-partite graph is on the nodes $\{\langle i, s\rangle \,|\, 1 \leq i \leq j \wedge s \in \Sigma^n\}$. The $j$ parts of the $j$-partite graph are the sets $\{\langle 1, s\rangle \,|\, s \in \Sigma^n\}$, $\{\langle 2, s\rangle \,|\, s \in \Sigma^n\}$, $\{\langle 3, s\rangle \,|\, s \in \Sigma^n\}$, etc. And given integers $1 \leq i < i' \leq j$ and strings $s, s' \in \Sigma^n$, there is a directed edge from $\langle i, s\rangle$ to $\langle i', s'\rangle$ if the circuit on input $0^{(i-1)(n+1)}1s0^{(i'-i-1)(n+1)}1s'0^{(j-i')(n+1)}$ evaluates to 1, and otherwise there is a directed edge from $\langle i', s'\rangle$ to $\langle i, s\rangle$. Let us explain this. We view the input as having $j$ $n$-bit fields, and just before each of those fields, a 1-bit control input. Basically, the $i$th field is about the $i$th part of the $j$-partite graph. When exactly two of the fields are activated via their control wires, and all the other inputs are 0, the circuit tells us about the edge between appropriate nodes (named by the parts and the input bits), which *per force* are from different parts. Note that this model ensures that any circuit (of the right input and output size/structure) specifies a $j$-tournament. Note also that, analogously to having always a power of two number of nodes in the Galperin–Wigderson model, in this variant of that for multipartite graphs, we always have that each part of the multipartite graph is of the same size as the others, and each of those parts is in cardinality a power of two.

We can now define the sets that will let us study the complexity of $k$-kingship in $j$-partite tournaments.

For each integer $k \geq 1$ and each integer $j \geq 2$, define the following set.

$(k, j)$-Tournament-Kings $= \{\langle c, x\rangle \,|\, c$ is a $j$-tournament-circuit and $x$ is a $k$-king in the $j$-partite tournament specified by $c\}$.

We now completely classify the complexity of $k$-kingship in $j$-partite tournaments.

**Theorem 4.1** *For each $k \geq 1$ and $j \geq 2$, $(k, j)$-Tournament-Kings is in P when $k = 1$ and is $\Pi_2^p$-complete otherwise.*

We prove this via the following collection of lemmas. It is immediately clear that these lemmas yield the result. (Regarding $\Pi_2^p$-hardness, this holds via double induction with (2,2) as the base case.)

**Lemma 4.2** *For each $j \geq 2$, $(1, j)$-Tournament-Kings is in P.*

**Lemma 4.3** *$(2, 2)$-Tournament-Kings is $\Pi_2^p$-hard.*

**Lemma 4.4** *For each $k \geq 1$ and $j \geq 2$, $(k, j)$-Tournament-Kings $\leq_m^p (k, j+1)$-Tournament-Kings.*

**Lemma 4.5** *For each $k \geq 1$, $(k, 2)$-Tournament-Kings $\leq_m^p (k+1, 2)$-Tournament-Kings.*

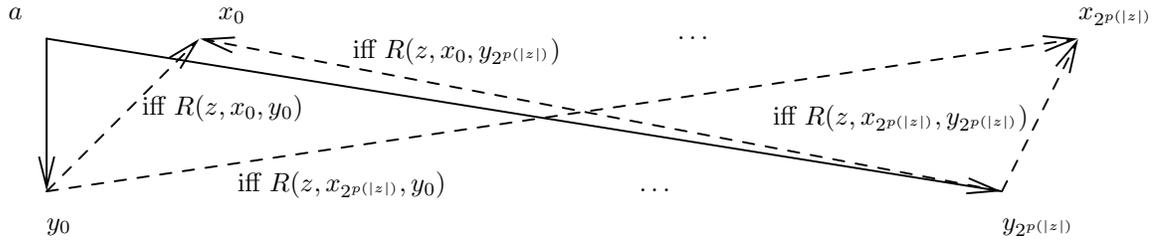

Figure 6: Figure for Lemma 4.3

**Lemma 4.6** *For each $k \geq 2$ and $j \geq 2$, $(k,j)$-Tournament-Kings $\in \Pi_2^p$.*

We now very briefly present the proofs of each of these lemmas. As noted earlier, easily handled issues of exact encodings and padding strings will not be mentioned.

**Proof of Lemma 4.2** Spiritually speaking, this should really be coNP-complete. In particular, the construction described in the text before Theorem 3.2, with all the cross-edges removed from between and within potential certificate nodes and padding nodes, *is* a bipartite tournament.

However, the fact that our model is such that each partition is of equal size causes this to be not coNP-complete but rather to be in P. This is simply because the only case when a 2-partite tournament (specified by a 2-tournament-circuit, and thus balanced) has a 1-king is when each part has one node, in which case the node that points to the other is a 1-king. If there is more than one node in a given part, we can have no 1-king since no node can reach any other node in its part via paths of length 1 (exactly because this *is* a bipartite tournament). The same holds not just in 2-partite tournaments but for $j$-partite tournaments, $j \geq 2$.

This is a fluke effect of $k = 1$. For $k > 1$, given a natural construction for an unbalanced multipartite tournament, it is not hard to see that one can complete it with dummy nodes without ruining the construction. But as we have just noted, for $k = 1$ this is not so. ❏

**Proof of Lemma 4.3** The most important facets of the construction from the first stage of the proof of Theorem 2.1 can be embedded into a 2-partite tournament. In particular, if we are trying (in light of some input $z$ for some arbitrary, fixed $\Pi_2^p$ set) to test that for all appropriate-length $x$ there exists an appropriate-length $y$ such that some P-time predicate $R(z, x, y)$ holds, we can do so via testing whether the node marked $a$ is a 2-king in the 2-partite tournament shown in Figure 6. Here we have not shown padding nodes, but it is clear how to handle those. Also, we have assumed, as is legal, that the $x$ and $y$ strings are of the same length as each other. ❏

**Proof of Lemma 4.4** Suppose we have a $(k, j)$ problem, with a designated node that we are interested in. We can easily tweak the $j$-tournament circuit to make it a $j + 1$-tournament circuit whose first $j$ parts correspond exactly to those of the original circuit, with the same edges, and such that the $j + 1$st part has the same number of nodes as each of the other parts, and all nodes between the $j + 1$st part and any of the other parts always point toward the $j+1$st part. It is clear that the designated node is a $k$-king in the original $j$-partite tournament exactly if it is a $k$-king in our new $j+1$-partite tournament. ☐

**Proof of Lemma 4.5** Let $k$ be a fixed integer greater than or equal to 1. Suppose we have a $(k,2)$ problem, with a designated node, $w$, whose $k$-kingship we are interested in. We can easily tweak the 2-tournament-circuit to make it a new 2-tournament-circuit whose parts exactly correspond to the previous circuit's nodes, except we add one new node, $z$, in the part of the tournament to which $w$ does not belong. $z$ points to $w$, but all other nodes in $w$'s part point to $z$. It is clear that $z$ is a $k+1$-king in the original 2-tournament exactly if $w$ is a $k$-king in the original 2-tournament. (As usual, we have not mentioned adding padding nodes to ensure that both parts have cardinality equal to the same power of two, but in this case that is easy. The padding nodes will be pointed to by all nodes from the other part, except between new padding nodes the edge can be put in in any way, e.g., always from part 1 to part 2.) ☐

**Proof of Lemma 4.6** This is immediately clear: For each node $v$, we guess a path of length at most $k$ from $w$ to $v$. ☐

The above completes our proof of Theorem 4.1.

All that remains is to make, as we promised earlier, some comments on the model that we chose not to use, namely, allowing our input to freely be any graph (specified succinctly via a circuit) and a node, and then asking the complexity of the question (for each fixed $k$ and $j$) "Is it the case that the node is a $k$-king in the specified graph and that the graph is a $j$-partite tournament?" Note that this puts into the set the burden not just of testing kingship but also of testing whether the graph is a $j$-partite tournament. (Here, we are not even trying to ensure that the parts are of the same size, as that seems to make things even harder—potentially requiring sequential exponential time.) However, we argue that, in this model, for each fixed $j > 1$, checking whether a succinctly specified Galperin-Wigderson-model graph is a $j$-partite tournament is relatively easy, namely, it is in coNP. To see this, we will need a characterization-by-excluded-subgraphs for $j$-partite tournaments. Throughout this paper, all graphs are assumed to be directed, but in the rest of this section we will need to speak of both directed and undirected graphs, so we will for the rest of this section be very explicit as to which we mean. Very relevant here are the following theorems. ($K_j$ denotes a simple, undirected $j$-clique.)

**Theorem 4.7** 1. *(see [Wes96, Exercise 1.3.37a] and [BvdHL04, p. 269]) A simple, undirected graph is a complete multipartite graph (i.e., there is a $j > 1$ such that it is a complete $j$-partite graph) if and only if $G$ is $(K_2 \cup K_1)$-free (i.e., it has no induced 3-node subgraph having exactly one edge).*

2. *(see [BvdHL04, p. 269]) A simple, undirected graph is a complete 2-partite graph if and only if $G$ is $(K_3, K_2 \cup K_1)$-free (i.e., it has no induced 3-node subgraph having exactly one or three edges).*

It is easy to see from the first part of the above result that the following characterization holds (in fact, the second part of the above result is the $j = 2$ case of this).

**Theorem 4.8** *Let $j > 1$ be fixed. A simple, undirected graph is a complete $j$-partite graph*

*if and only if $G$ is $(K_{j+1}, K_2 \cup K_1)$-free.*[3]

Now, those results are for simple undirected graphs, but those results easily yield analogous results for simple directed graphs, and in particular we have the following result. Here, $Q_2$ denotes the 2-node simple directed graph in which each node points to the other node. And given any simple, directed graph $G = (V, E)$, as usual the *underlying graph* of $G$, denoted here $Under(G)$, is the simple, undirected graph whose node set is $V$ and whose edge set is $\{\{a, b\} \mid (a, b) \in E\}$.

**Theorem 4.9** *Let $j > 1$ be fixed. A simple, directed graph $G$ is a $j$-partite tournament if and only if $G$ is $Q_2$-free and $Under(G)$ is $(K_{j+1}, K_2 \cup K_1)$-free.*

Theorem 4.9 makes it immediately clear that, for each fixed $j > 1$, testing whether a succinctly specified (in the Galperin–Wigderson model) graph is a $j$-partite tournament is a coNP test, since the theorem exactly makes it so. Stated more explicitly, we have the following.

**Theorem 4.10** *Let $j > 1$. The set $\{c \mid$ the graph specified (in the Galperin–Wigderson model) by circuit $c$ is a $j$-partite tournament$\}$ belongs to coNP.*

And in light of Theorem 4.9 and in particular the just-stated result, it is not hard to see that (though one has to as always be careful regarding padding nodes—a $j$-partite tournament in our family model has $j$ times a power of two nodes, and each part is of equal size, but in the Galperin–Wigderson model and the result we are about to mention, parts can differ in size, and the total graph is of size a power of two; however, the only real worry about checking complete $j$-partite-ness was whether the upper bound would prove a problem, and the coNP-result above removes that as a worry), for each fixed $j > 1$, our general techniques easily yield that the set

$\{\langle c, x \rangle \mid c$ has $2|x|$ inputs and $x$ is a king in the graph specified by $c$ and the graph specified by $c$ is a $j$-partite tournament$\}$,

(which is the natural set to study in the model where one does require the circuit to itself enforce "$j$-partite tournament"-ness) is $\Pi_2^p$-complete.

---

[3]For completeness, let us quickly prove this. Regarding the "only if" direction: If $G$ is a simple, undirected, complete $j$-partite graph, then by part 1 of Theorem 4.7 it must be $(K_2 \cup K_1)$-free, and clearly it is $K_{j+1}$-free, since if it has $K_{j+1}$ as an induced subgraph that immediately blocks $j$-partite-ness. Regarding the "if" direction of the theorem: Let $G$ be a simple, undirected graph that is $(K_{j+1}, K_2 \cup K_1)$-free. By part 1 of Theorem 4.7, the $(K_2 \cup K_1)$-free-ness implies that for some $j' > 1$ it holds that $G$ is complete $j'$-partite. But if the smallest such $j'$ is $j + 1$ or greater, then each of the $j'$ parts must be nonempty (since otherwise $j'$ would *not* be the smallest such value). However, those parts being nonempty means the graph would have $K_{j'}$ as an induced subgraph (namely, the one induced from choosing one node from each of the $j'$ parts—since $j'$ is a value for which it is a *complete $j'$-partite graph*), which contradicts the assumption that it is $K_{j+1}$-free. Thus, $G$ is certainly a complete $j$-partite subgraph.

## 5   Conclusions and Open Problems

We have seen that for some succinctly specified families of tournaments, the king problem is $\Pi_2^p$-complete. We also showed that for succinctly specified general graphs the king problem is $\Pi_2^p$-complete, and we explored other complexities of king problems, how associativity affects such complexity, and what holds for $k$-kings in tournaments and $j$-partite tournaments.

For a set $A$, the $\equiv_m^p$-degree (polynomial-time many-one equivalence degree) of $A$ is $\{L \mid L \leq_m^p A \wedge A \leq_m^p L\}$. One question that might be natural for additional study is to classify which $\equiv_m^p$-degrees pinpoint the complexity of the king problem of some tournament family specifier. For example, we already showed that the $\equiv_m^p$-degrees of the complete problems for $\Pi_2^p$, coNP, and NP contain the king problems of some tournament family specifiers. And note that the extremely computationally simple tournament family specifier $f(x,y) = \max(x,y)$ is such that $\text{Kings}_f$ belongs to the $\equiv_m^p$-degree made up of the sets $\text{P} - \{\emptyset, \Sigma^*\}$. Also, it is very easy to see that the $\equiv_m^p$-degrees of $\emptyset$ and $\Sigma^*$ do not contain the king problem of any tournament family specifier. Can one show—unconditionally, as this easily holds if one assumes P = NP—that every $\equiv_m^p$-degree contained in $\Pi_2^p$, other than those two, contains some tournament family specifier's king set?

Also, Theorem 2.1 indeed shows that the $\Pi_2^p$ upper bound of [HOZZ] is optimal. But regarding the $\Pi_2^p/1$-nonimmunity upper bound of Theorem 1.2, the result of Theorem 2.1 merely shows that one particular attack on that claim's optimality is unlikely to succeed. Can one provide more direct evidence of the optimality of Theorem 1.2? For example, can one show that if each infinite P-selective set has an infinite $\Sigma_2^p/1$ (i.e., $\text{NP}^{\text{coNP}}/1$) subset, then some unexpected complexity class equality or hierarchy collapse follows?

Finally, our study of multipartite tournaments was in the model of individually specified graphs. What results hold for an analog of family specifiers in that setting?